\documentclass[9pt,twocolumn,twoside]{opticajnl}
\journal{opticajournal} % use for journal or Optica Open submissions

\setboolean{shortarticle}{true}
\usepackage{lineno}
\title{Photonic Extreme Learning Machines using Event-Based detection}

\author[1,2,*]{Vicente Rocha}
\author[1,2]{Tomás Lopes}
\author[1,2]{Joana Teixeira}
\author[1]{Tiago Ferreira}
\author[1]{Catarina Monteiro}
\author[1,2]{Nuno Silva}

\affil[1]{INESC TEC, Centre of Applied Photonics, Rua do Campo Alegre 687, 4169-007 Porto, Portugal}
\affil[2]{Faculdade de Ciências da Universidade do Porto, Departmento de Física e Astronomia, Rua do Campo Alegre 687, 4169-007 Porto, Portugal}

\affil[*]{vicente.v.rocha@inesctec.pt}

\begin{abstract}

Photonic extreme learning machines use random optical propagation, detection nonlinearity, and a trained linear readout for energy-efficient and scalable optical computing. However, conventional intensity readout with CCD or CMOS cameras constrain the dimensionality of the hidden representation space. Here, we experimentally replace intensity detection with event-based camera, whose thresholded log-intensity response provides alternative pixel-wise hidden representations: first-event time, binary activation, and event count. In nonlinear two spiral classification task we obtain accuracies of $93\pm 3\%$ with strong intrinsic generalization, and comparable ridge and pseudo-inverse performance indicating sample-limited effective dimensionality. Regression results reveal sensitivity to systematic optical-intensity drift, identifying stability requirements for future event-based PELMs. These results establish event-based detection as a route toward richer photonic hidden representations while clarifying current limitations.

\end{abstract}

\setboolean{displaycopyright}{false} % Do not include copyright or licensing information in submission.

\begin{document}

\maketitle

\section{\label{sec: introduction} Introduction}

    Photonic Extreme Learning Machines (PELM) have recently emerged as energy-efficient neuromorphic processors bypassing training of the hidden layer, enabling scalable optical learning architectures~\cite{pierangeli2021photonic}. Their versatility has led to multiple implementations including integrated circuits~\cite{biasi2023array, rausell2025programmable}, optical fibers~\cite{redding2024fiber}, and free-space propagation~\cite{pierangeli2021photonic}. However, their performance can be limited by the dimensionality of the hidden space, which is constrained by the available optical nonlinearities and detector response functions~\cite{rocha2025probing}. Recent works on PELMs for Natural Language Models~\cite{m2022large}, camera saturation responses~\cite{pierangeli2021photonic, rocha2025probing}, and detection with perovskite-solar-cells~\cite{zhang2025optical} motivate the search for alternative readouts for higher-performance machines.

    At the foundational level, PELMs build on the Extreme Learning Machine (ELM) framework~\cite{huang2004extreme, li2005fully}, consisting of a random hidden layer with non-polynomial activation functions, followed by a trained linear output layer which is capable of universal function approximation and features intrinsic generalization capabilities~\cite{huang2006extreme}. A direct mapping between the optical systems and ELM framework was established in~\cite{pierangeli2021photonic, rocha2025probing}. In PELMs, speckle generation provides the random linear weights and bias, while the interferometric process and saturated camera response define a nonlinear activation function~\cite{pierangeli2021photonic,rocha2025probing}. However, due to the properties of the activation function related to the square-law detection, the intrinsic generalization capabilities of PELMs with a saturated detector are hardly observed if the input space is of low dimensionality, in particular due to the use of additional regularizations filtering the dimensions carrying noise~\cite{rocha2025probing}.

    Motivated by this context and the recent emergence of event-based camera detectors for optical sensing applications~\cite{gallego2020event, guo2024eventlfm, ni2012asynchronous, willert2022event, su2025inter}, this work explores the use of this type of detector to generate distinct pixel-wise nonlinear activation functions based on the native thresholded log-intensity variation of such detectors. In particular, in section~\ref{sec: Results} we evaluate the resulting Event-based PELMs (EV-PELM) on classification and regression tasks using ridge and pseudo-inverse linear readouts, and relate the results with the increase of the usable rank of the hidden-layer output matrix, estimated from regularization dependence and effective dimensionality. Generalization capability is also assessed under repeated noisy measurements, fueling a final discussion in Section~\ref{sec: discussion} regarding the opportunities and limitations of event-based detection for photonic neuromorphic computing in ELM architectures.

\begin{figure*}%[ht]
\centering
\includegraphics[width=1.0\textwidth]{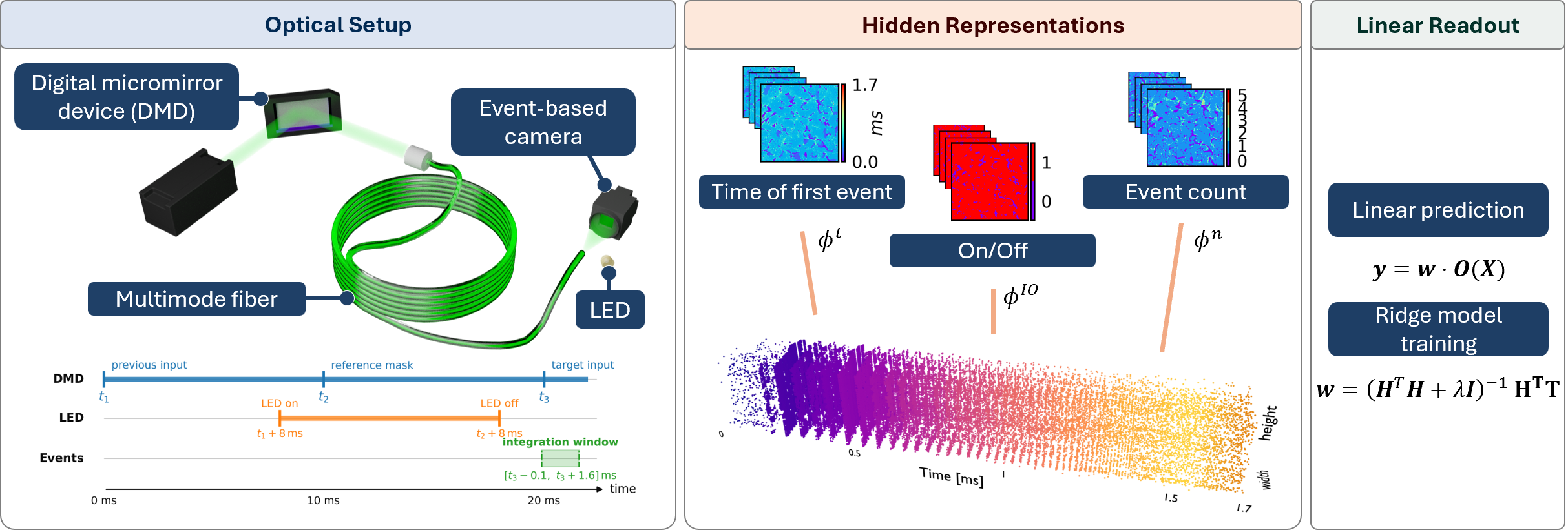}
\caption{Event-based photonic extreme learning machine. Input
features are encoded on a digital micromirror device and
propagated through a multimode fiber, whose random transmission
implements the fixed hidden-layer projection. The resulting
speckle field is detected by an event-based camera. Each pixel is
then mapped into one of three hidden representations: time of first
event, binary activation, or total event count. The final layer is a
digital linear readout obtained analytically.}
\label{fig:setup}
\end{figure*}

%\begin{figure*}[ht]
%    \centering
%    \includegraphics[width=1.\textwidth]{Figures//setup.png}
%    \caption{Event-based photonic extreme learning machine. Input features are encoded on a digital micromirror device and propagated through a multimode fiber, whose random transmission matrix implements the fixed hidden-layer weights. The resulting speckle field is detected by an event-based camera, and each pixel is mapped into one of three hidden representations: time of first event, binary On/Off activation, or total event count. The final layer is the digital linear regression featuring an analytic solution.}
%    \label{fig: setup}
%\end{figure*}

\section{Mathematical formalism and operation principles}

    An ELM maps each input vector $\boldsymbol{X}_i$ to a high-dimensional hidden representation $\boldsymbol{O}\left(\boldsymbol{X}_i\right)$, followed by a trained linear readout $\boldsymbol{\beta}$ yielding a prediction $\boldsymbol{y}(\boldsymbol{X}_i)=\boldsymbol{\beta}\odot \boldsymbol{O}(\boldsymbol{X}_i)$\cite{huang2006extreme}. In a photonic ELM, the information is typically encoded in the amplitude/phase profile of the input optical beam (e.g. an image), before undergoing a random projection implemented optically by propagation through a complex medium, followed by a square-law detection that provides the effective nonlinear activation. For $N_{\mathrm{train}}$ samples and $N_p$ detector pixels, the hidden-layer output matrix is
\begin{equation}
    \bar{\boldsymbol{H}}=
    \begin{bmatrix}
        O_1(\boldsymbol{X}_1) & \cdots & O_{N_p}(\boldsymbol{X}_1) \\
        \vdots & \ddots & \vdots \\
        O_1(\boldsymbol{X}_{train}) & \cdots & O_{N_p}(\boldsymbol{X}_{train})
    \end{bmatrix},
    \label{eq:hidden_matrix}
\end{equation}
    Under the theory of the ELMs if one can write $\boldsymbol{O}(\boldsymbol{X}_i)=G(\bar{\boldsymbol{W}}\boldsymbol{X}_i+\boldsymbol{b})$ it can be shown that if the weights $\bar{\boldsymbol{W}}$ are extracted from a random distribution, and $G$ is a non-polynomial nonlinear activation function, $\bar{\boldsymbol{H}}$ will be full-rank and the ELM will feature universal approximation capabilities~\cite{huang2004extreme,huang2006extreme}.

    To obtain the output weights of the linear readout layer $\boldsymbol{\beta}$, multiple methods can be used depending on the task and regularization~\cite{hastie2009elements}. A robust method is the ridge regression, that for a training $\boldsymbol{\bar{H}}$ and associated target matrix $\boldsymbol{\bar{T}}$, gives a solution for the readout layer as
\begin{equation}
    \boldsymbol{\beta}_{\lambda}
    =
    \left(\boldsymbol{H}^{T}\boldsymbol{H}
    + \lambda \boldsymbol{I}\right)^{-1}
    \boldsymbol{H}^{T}\boldsymbol{T},
    \label{eq:ridge_solution}
\end{equation}
    where $\lambda$ controls the regularization strength, with $\lambda=0$ the Moore-Penrose pseudo-inverse solution. The role of $\lambda$ can be understood from the singular values $\sigma_i$ of $\boldsymbol{\bar{H}}$. Indeed, the effective dimensionality for a ridge regression is a common metric for the flexibility of the model and is defined as~\cite{hastie2009elements}
\begin{equation}
    d_{\mathrm{eff}}
    =
    \sum_{i=1}^{\min(N_{train},N_p)}
    \frac{\sigma_i^2}{\sigma_i^2+\lambda}.
    \label{eq:effective_dimensionality}
\end{equation}
    Thus, ridge regression suppresses directions with $\sigma_i^2 \lesssim \lambda$ and preserves larger components. In the context of PELMs, this suppression can mitigate the experimental noise~\cite{rocha2025probing} but effectively it reduces the dimensionality of the hidden representation space. Scanning $\lambda$ therefore probes directly the usable dimensionality of the hidden representation.

\begin{figure*}[!t]
    \centering
    \includegraphics[width=.7\textwidth]{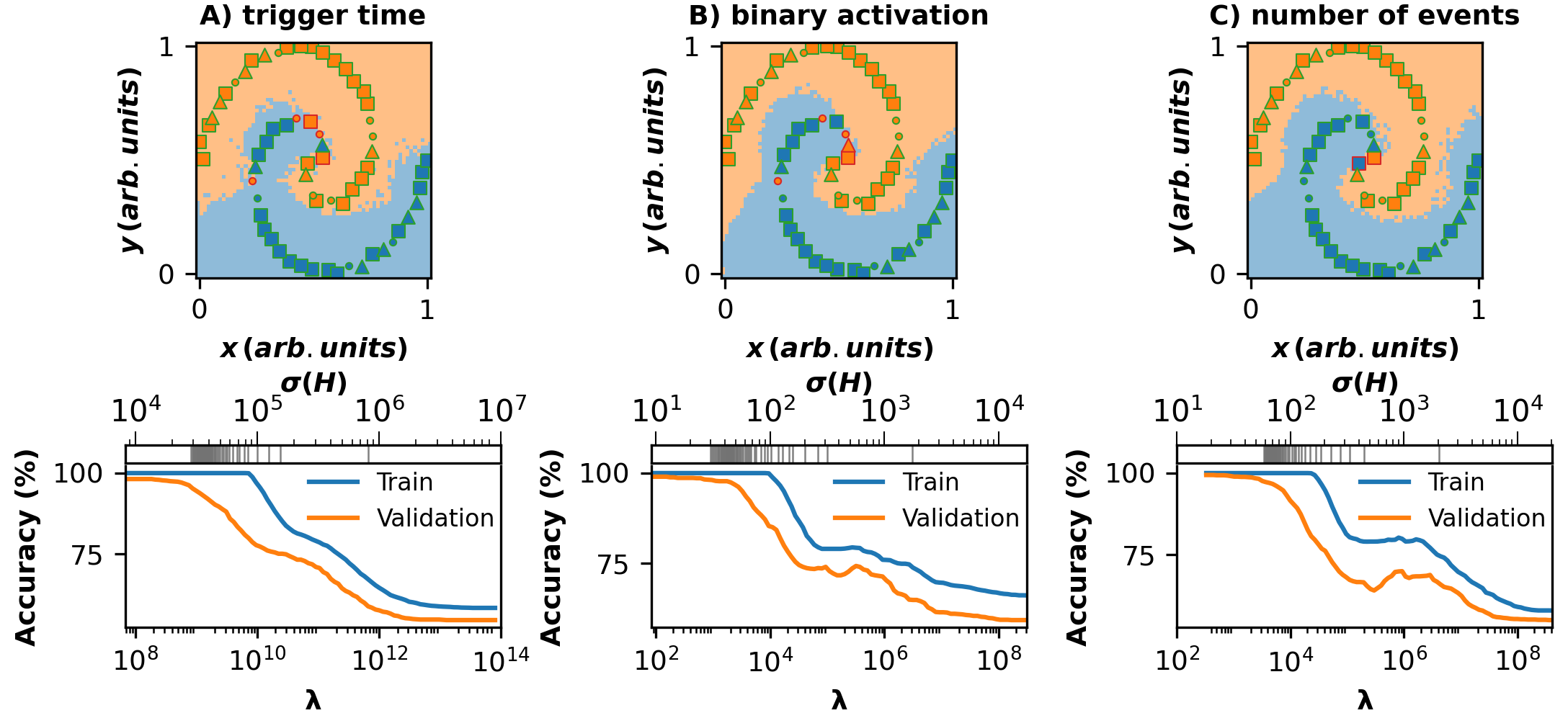}
    \caption{Spiral classification using the three event-based hidden representations. Top row: decision maps obtained from a trained linear readout with test sample overlaid. Shapes indicate the subset - Squares/dots/triangles correspond to training/validation/tests points, respectively - whereas edges correspond to correct (green) or incorrect (red) prediction. Bottom row: train and validation accuracy as a function of the ridge parameter $\lambda$, as detailed in the main text.}
    \label{fig: classification results}
\end{figure*}

    Event-based PELMs, as represented in Figure~\ref{fig:setup}, replace conventional intensity readout with asynchronous event detection, providing a nonlinear and sparse optical-to-digital mapping with potentially lower energy consumption~\cite{gallego2020event}. Event detection with dynamic vision sensors (DVS) records binary events and their timestamps whenever the log intensity at a pixel changes by a threshold value. Combining the PELM model (see \cite{rocha2025probing} for details) with the DVS response~\cite{gallego2020event}, the pixel-wise log-intensity can be written as
\begin{equation}
    L\left(\boldsymbol{X}\right) = \log\left(|\bar{\boldsymbol{w}} \cdot \boldsymbol{E}^{\left(in\right)}\left(X\right) + \boldsymbol{b}|^2 + \boldsymbol{B}\right),
\end{equation}
    where $\boldsymbol{E}^{\left(in\right)}\left(\boldsymbol{X}\right)$ is the input state whose transverse mode is amplitude-modulated with the digital micromirror device (DMD) to encode the feature $\boldsymbol{X}$ (see Figure \ref{fig:setup}), $\bar{\boldsymbol{w}}$ and $\boldsymbol{b}$ are the random linear weights and bias provided by the transmission on the complex medium and unmodulated part of the beam, and the logarithmic intensity response defines the scale for the detection-induced activation function with $\boldsymbol{B}$ denoting the background signal, ensuring the equation remains defined in the absence of the beam. An event is registered at pixel $\boldsymbol{r}_k=(x_k,y_k)$ when
\begin{equation}
    \Delta L\left(\boldsymbol{r}_k, t_k\right) = |L\left(\boldsymbol{r}_k, t_k\right) - L\left(\boldsymbol{r}_k, t_{k-1}\right)| > C,
\end{equation}
    with $C$ being a tunable event threshold and $t_{k-1}$ the previous event time. The event can also have a polarity $p=\pm1$ depending if it corresponds to an increase or decrease in the pixel intensity. During each acquisition, the detector returns a set of events
$\mathcal{E}=\{e_k\}_{k=1}^{N}$, containing pixel addresses,
timestamps, and polarities. This event stream may be converted into the
ELM hidden state through a pixel-wise mapping as
\begin{equation}
    \boldsymbol{O}(\boldsymbol{X})
    =
    \phi\left(\mathcal{E}\right),
    \label{eq:outputs}
\end{equation}
    where $\phi$ defines the detector-induced activation. We test three pixel-wise mappings $\phi$: A. the time of the first event at each pixel, B. a binary activation indicating whether a pixel fired, and C. the total number of events recorded by each pixel. These thresholded responses can enrich the hidden space beyond polynomial square-law activation, but may also inherit DVS nonidealities such as missed events at high rates, charge accumulation, and noise-triggered events~\cite{mcreynolds2022experimental}. Therefore, a careful evaluation of EV-PELMs in terms of noise robustness, representation dimensionality, and generalization, in specific for cases of low dimensionality of the input feature space, is highly relevant for optical computing and in specific PELMs.

%%%%%%%%%%%%%%%%%%%%%%%%%%%%%%%%%%
%%%%%%%%%%%%%%%%%%%%%%%%%%%%%%%%%%

\section{\label{sec: Results} Experimental Results and Model Validation}

    To evaluate the EV-PELM against typical intensity-based PELMs, we focused on nonlinear low-dimensional tasks where PELMs are known to struggle~\cite{rocha2025probing}. For this we will evaluate the EV-PELM on two standard yet highly challenging benchmark tasks for this type of hardware: classification of two spiral distributions and regression of a target sinc function. For the classification another two distributions were tested and are presented in the supplementary material Section 1.A.

    The acquisition protocol used to induce reproducible feature-dependent events is shown in Figure~\ref{fig:setup}. The event-based camera records continuously, requiring synchronization between the DMD, background LED, and the camera. Input features $\boldsymbol{X}$ are encoded on the optical beam with the DMD using Jarvis-Judice-Ninke dithering to approximate continuous amplitudes with binary micromirror states~\cite{jarvis1976survey}. The DMD updates every $10\, \mathrm{ms}$, cycles through previous input, reference mask, and target input at $t_1$, $t_2=t_1 + 10\, \mathrm{ms}$, and $t_3=t_1+20\,\mathrm{ms}$, respectively. Delayed LED switching at $t_2 + 8\,\mathrm{ms}$ erases previous event history and sets the reference state. Feature-dependent events are integrated over $\left[t_3 - 0.1\, \mathrm{ms}, \, t_3+1.6\,\mathbf{ms}\right]$. Each dataset is acquired by iterating this protocol over the set of input features, while the full dataset is composed of 30 repetitions of all feature measurements resulting in 30 distinct samplings of noise under the same input features.

    Starting our analysis with the spiral classification task, Figure~\ref{fig: classification results} shows good prediction accuracy and efficient learning of the true spiral distribution on the background decision maps for the three pixel-wise features. Furthermore, the accuracy curves show the best performance for weak or negligible additional regularization in the ridge model.

\begin{figure*}[ht]
    \centering
    \includegraphics[width=.7\textwidth]{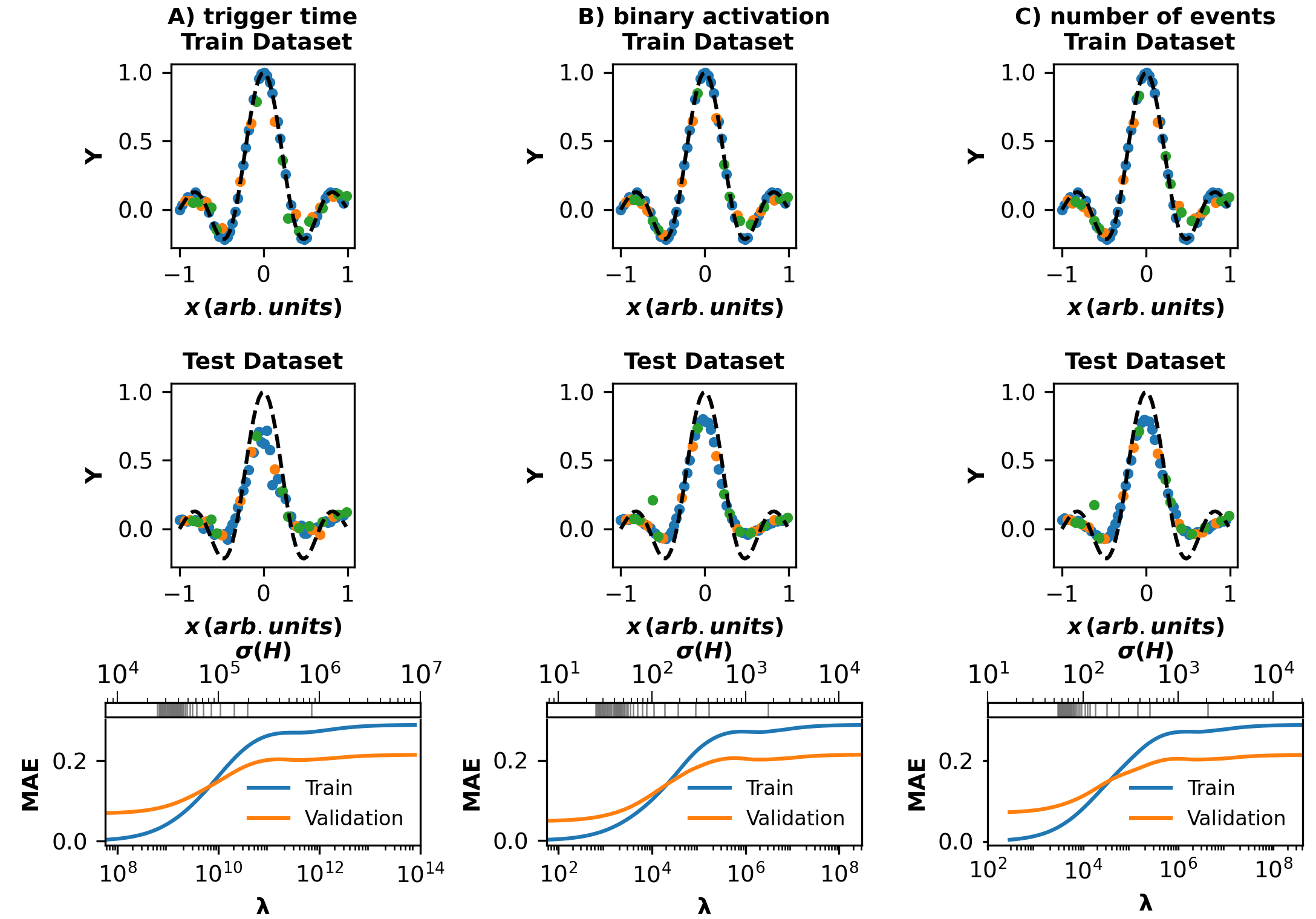}
    \caption{Sinc regression using the three event-based hidden representations. Top and middle row shows predictions within the train and test datasets for the train (blue), validation (orange), and test (green) points. The dashed black curve corresponds to the target sinc function. Bottom row shows train and validation mean absolute error curves as function of the ridge parameter $\lambda$.}
    \label{fig: regression results}
\end{figure*}

    The regression task reveals the main limitation of the configuration. As shown in Figure~\ref{fig: regression results}, predictions for the train dataset accurately follow the target sinc function, including points not used for fitting (validation and test points). However, on independent test repetitions (e.g. distinct single run), the prediction trend remains visible but the error increases, indicating representation drift rather than simple additive noise. This drift may be related with intensity fluctuations, and the associated DVS response to optical-intensity fluctuations. For example a change in laser intensity shifts the time required for each pixel to reach the event threshold, modifying the first-trigger map. Besides, it can also turn pixels on or off and change the total number of events. Thus, slow intensity variations can systematically alter the hidden representation between repetitions. In line with the curves for classification, the MAE obtained shows best generalization performance for small additional regularization.

    Quantitatively, the $5$-fold cross-validation on independent test points of test repetitions gives the performance summarized in Table S1 and Table S2 for classification and sinc regression, respectively (supplementary material). For spiral classification, the pseudo-inverse achieves mean accuracy of $83\%$, $89\%$, and $93\%$ for first-trigger time, binary activation, and event count, respectively. For sinc regression, the corresponding MAEs are $0.13$, $0.10$, and $0.11$. The classification results correspond to higher performances than previously obtained for saturation activation functions, but regression errors may perform worse in this framework~\cite{rocha2025probing}. Finally, we also note that the comparable performance between ridge and the pseudo-inverse solution indicates the rank of $\boldsymbol{H}$ reaches the sample limit. Effectively, we can make $\lambda=0$, from which equation~\ref{eq:effective_dimensionality} gives $d_{eff}=\min(N_f,N_p)$ for nonzero singular values. Since $N_p \gg N_f$, the hidden representation is limited by the number of input samples, yielding $d_{eff}=N_f=64$.

%All of the results correspond to a higher performance than that previously obtained for saturation activation function for the same problem~\cite{rocha2025probing}, which validates the advantage of this type of detector in the context of PELMs. Finally, we also note that the comparable performance of ridge regression and the pseudo-inverse solution indicates that the rank of $\boldsymbol{H}$ reaches the sample limit. Effectively, we can make $\lambda=0$, from which equation~\ref{eq:effective_dimensionality} gives $d_{eff}=\min(N_f,N_p)$ for nonzero singular values. Since $N_p \gg N_f$, the hidden representation is limited by the number of input samples, yielding $d_{eff}=N_f=64$.

%\begin{table*}[h]
%    \centering
%    \begin{tabular}{c | c | c | c | c | c | c |}
%        \cline{2-7}
%         &  \multicolumn{3}{| c |}{Spiral} & \multicolumn{3}{| c |}{Sinc}  \\
%         \cline{2-7}
%         &  \multicolumn{3}{| c |}{Accuracy ($\%$)} & \multicolumn{3}{| c |}{MAE}  \\
%         \cline{2-7}
%         & trigger time  & On/Off & total events & trigger time  & On/Off & total events \\
%         \cline{1-7}
%         \multicolumn{1}{| c |}{pseudo-inverse} & $83\pm 6$ & $89\pm 5$ & $93\pm 3$ & $0.13\pm 0.03$ & $0.10\pm 0.03$ & $0.11\pm 0.02$  \\
%         \cline{1-7}
%         \multicolumn{1}{| c |}{Ridge} & $83\pm 5$ & $88\pm 5$ & $93\pm 3$ & $0.13\pm 0.03$ & $0.11\pm 0.03$ & $0.11\pm 0.02$  \\
%         \cline{1-7}
%         \multicolumn{1}{| c |}{effective rank}  & 64 & 64 & 64 & 64 & 64 & 64  \\
%         \cline{1-7}
%    \end{tabular}
%    \caption{Cross-validated performance of the event-based PELM for spiral classification and sinc regression. Values report the mean and standard deviation over the evaluated folds and independent test repetitions. Pseudo-inverse and ridge readouts achieve comparable performance, indicating that the useful hidden-space dimensionality is limited primarily by the number of input samples rather than by the number of camera pixels. The effective rank estimation is obtained from equation~\ref{eq:effective_dimensionality}. In all the cases, the pseudo-inverse achieves better results compared with ridge regression so the best model occurs for $\lambda=0$ yielding a $64$ dimensionality.}
%    \label{tab: results}
%\end{table*}

\section{\label{sec: discussion} Discussion and Concluding Remarks}

    Event-based cameras provides an alternative optical-to-digital nonlinear response based on intensity variations rather than absolute intensity. In the context of PELMs, this response increases the dimensionality of the hidden representation while preserving the simplicity of a trained linear output layer. However, as observed this information readout not only presents opportunities but also some challenges.

    Overall, and compared with the camera saturation-based PELM in~\cite{rocha2025probing} in low-dimensional feature spaces, the DVS readout achieves higher classification accuracy and stronger generalization in spiral task. This is consistent with the higher nonlinearity of the detector response compared with the saturable response of a standard camera, allowing a higher effective dimensionality of the intermediate space. Indeed, the comparable performance of ridge and pseudo-inverse readouts in both tasks indicates that the hidden representation reaches the sample-limited rank, which was a problem previously identified in saturation-based PELMs. This is due in part to the highly nonlinear readout introduced by the system, but also to the robustness provided by the event-generation threshold against small intensity fluctuations.

    However, further tests on the regression task also revealed a drift in the representation due to systematic fluctuations that modify event timing, binary activations, and event counts. While this effect is not very pronounced for classification it limits the system for regression tasks. Thus heading towards more general implementations, stabilizing the optical power, or adding calibration frames shall be considered to improve stability. Nevertheless, the results enclosed clearly present and map the capabilities of event-based PELMs over previously explored configurations, thus paving the way for future applications of such hardware solutions for fast and energy-efficient computational tasks.

\section{Back matter}

\begin{backmatter}

\bmsection{Acknowledgment} This work is financed by National Funds through the FCT - Fundação para a Ciência e a Tecnologia, I.P. (Portuguese Foundation for Science and Technology) within the project QuAC, with reference 2024.16086.PEX (https://doi.org/10.54499/2024.16086.PEX). V.R, T.L, and J.T. are supported by Fundação para a Ciência e Tecnologia through Grants No. 2025.07060.BD, 2024.01830.BD, and 2024.00426.BD, respectively.

\bmsection{Disclosures} The authors declare no conflicts of interest.

\bmsection{Data availability} Data underlying the results presented in this paper are not publicly available at this time but may be obtained from the authors upon reasonable request.

\bmsection{Supplemental document}
See Supplement 1 for supporting content.

\end{backmatter}

% Bibliography
\bibliography{sample}

@article{pierangeli2021photonic,
  title={Photonic extreme learning machine by free-space optical propagation},
  author={Pierangeli, Davide and Marcucci, Giulia and Conti, Claudio},
  journal={Photonics Research},
  volume={9},
  number={8},
  pages={1446--1454},
  year={2021},
  publisher={Chinese Laser Press and Optical Society of America}
}

@article{m2022large,
  title={Large-scale photonic natural language processing},
  author={M. Valensise, Carlo and Grecco, Ivana and Pierangeli, Davide and Conti, Claudio},
  journal={Photonics Research},
  volume={10},
  number={12},
  pages={2846--2853},
  year={2022},
  publisher={Chinese Laser Press and Optica Publishing Group}
}

@article{redding2024fiber,
  title={Fiber optic computing using distributed feedback},
  author={Redding, Brandon and Murray, Joseph B and Hart, Joseph D and Zhu, Zheyuan and Pang, Shuo S and Sarma, Raktim},
  journal={Communications Physics},
  volume={7},
  number={1},
  pages={75},
  year={2024},
  publisher={Nature Publishing Group UK London}
}

@inproceedings{huang2004extreme,
  title={Extreme learning machine: a new learning scheme of feedforward neural networks},
  author={Huang, Guang-Bin and Zhu, Qin-Yu and Siew, Chee-Kheong},
  booktitle={2004 IEEE international joint conference on neural networks (IEEE Cat. No. 04CH37541)},
  volume={2},
  pages={985--990},
  year={2004},
  organization={Ieee}
}

@article{huang2006extreme,
  title={Extreme learning machine: theory and applications},
  author={Huang, Guang-Bin and Zhu, Qin-Yu and Siew, Chee-Kheong},
  journal={Neurocomputing},
  volume={70},
  number={1-3},
  pages={489--501},
  year={2006},
  publisher={Elsevier}
}

@article{li2005fully,
  title={Fully complex extreme learning machine},
  author={Li, Ming-Bin and Huang, Guang-Bin and Saratchandran, Paramasivan and Sundararajan, Narasimhan},
  journal={Neurocomputing},
  volume={68},
  pages={306--314},
  year={2005},
  publisher={Elsevier}
}

@article{biasi2023array,
  title={An array of microresonators as a photonic extreme learning machine},
  author={Biasi, Stefano and Franchi, Riccardo and Cerini, Lorenzo and Pavesi, Lorenzo},
  journal={APL photonics},
  volume={8},
  number={9},
  year={2023},
  publisher={AIP Publishing}
}

@article{rausell2025programmable,
  title={Programmable photonic extreme learning machines},
  author={Rausell-Campo, Jos{\'e} Roberto and Hurtado, Antonio and P{\'e}rez-L{\'o}pez, Daniel and Capmany Francoy, Jos{\'e}},
  journal={Laser \& Photonics Reviews},
  volume={19},
  number={9},
  pages={2400870},
  year={2025},
  publisher={Wiley Online Library}
}

@article{zhang2025optical,
  title={Optical neural networks based on perovskite solar cells},
  author={Zhang, Kaicheng and Harwell, Jonathon and Pierangeli, Davide and Conti, Claudio and Di Falco, Andrea},
  journal={Photonics Research},
  volume={13},
  number={2},
  pages={382--386},
  year={2025},
  publisher={Chinese Laser Press and Optica Publishing Group}
}

@article{rocha2025probing,
  title={Probing a theoretical framework for a Photonic Extreme Learning Machine},
  author={Rocha, Vicente Vieira and Silva, Duarte and Moreira, Felipe Coelho and Monteiro, Catarina and Ferreira, Tiago David and Azevedo Silva, Nuno},
  journal={New Journal of Physics},
  year={2025}
}

@article{gallego2020event,
  title={Event-based vision: A survey},
  author={Gallego, Guillermo and Delbr{\"u}ck, Tobi and Orchard, Garrick and Bartolozzi, Chiara and Taba, Brian and Censi, Andrea and Leutenegger, Stefan and Davison, Andrew J and Conradt, J{\"o}rg and Daniilidis, Kostas and others},
  journal={IEEE transactions on pattern analysis and machine intelligence},
  volume={44},
  number={1},
  pages={154--180},
  year={2020},
  publisher={IEEE}
}

@article{mcreynolds2022experimental,
  title={Experimental methods to predict dynamic vision sensor event camera performance},
  author={McReynolds, Brian and Graca, Rui and Delbruck, Tobi},
  journal={Optical Engineering},
  volume={61},
  number={7},
  pages={074103--074103},
  year={2022},
  publisher={Society of Photo-Optical Instrumentation Engineers}
}

@misc{hastie2009elements,
  title={The elements of statistical learning},
  author={Hastie, Trevor and Tibshirani, Robert and Friedman, Jerome and others},
  year={2009},
  publisher={Springer series in statistics New-York}
}

@article{jarvis1976survey,
  title={A survey of techniques for the display of continuous tone pictures on bilevel displays},
  author={Jarvis, John F and Judice, Charles N and Ninke, William H},
  journal={Computer graphics and image processing},
  volume={5},
  number={1},
  pages={13--40},
  year={1976},
  publisher={Elsevier}
}

@article{guo2024eventlfm,
  title={EventLFM: event camera integrated Fourier light field microscopy for ultrafast 3D imaging},
  author={Guo, Ruipeng and Yang, Qianwan and Chang, Andrew S and Hu, Guorong and Greene, Joseph and Gabel, Christopher V and You, Sixian and Tian, Lei},
  journal={Light: Science \& Applications},
  volume={13},
  number={1},
  pages={144},
  year={2024},
  publisher={Nature Publishing Group UK London}
}

@article{ni2012asynchronous,
  title={Asynchronous event-based high speed vision for microparticle tracking},
  author={Ni, Zhenjiang and Pacoret, C{\'e}cile and Benosman, Ryad and Ieng, Siohoi and R{\'E}GNIER*, St{\'e}phane},
  journal={Journal of microscopy},
  volume={245},
  number={3},
  pages={236--244},
  year={2012},
  publisher={Wiley Online Library}
}

@article{willert2022event,
  title={Event-based imaging velocimetry: an assessment of event-based cameras for the measurement of fluid flows},
  author={Willert, Christian E and Klinner, Joachim},
  journal={Experiments in Fluids},
  volume={63},
  number={6},
  pages={101},
  year={2022},
  publisher={Springer}
}

@article{su2025inter,
  title={Inter-event interval microscopy for event cameras},
  author={Su, Changqing and Chen, Yanqin and Lin, Zihan and Cheng, Zhen and Zhou, You and Xiong, Bo and Yu, Zhaofei and Huang, Tiejun},
  journal={Photonics Research},
  volume={13},
  number={10},
  pages={2843--2853},
  year={2025},
  publisher={Chinese Laser Press and Optica Publishing Group}
}

% Full bibliography added automatically for Optics Letters submissions; the following line will simply be ignored if submitting to other journals.
% Note that this extra page will not count against page length

\bibliographyfullrefs{sample}

%Manual citation list
%\begin{thebibliography}{1}
%\bibitem{Zhang:14}
%Y.~Zhang, S.~Qiao, L.~Sun, Q.~W. Shi, W.~Huang, %L.~Li, and Z.~Yang,
 % \enquote{Photoinduced active terahertz metamaterials with nanostructured
  %vanadium dioxide film deposited by sol-gel method,} Opt. Express \textbf{22},
  %11070--11078 (2014).
%\end{thebibliography}

\end{document}